# Observations of interior whispering gallery modes in asymmetric optical resonators with rational caustics

Jie Gao[*], Pascal Heider[*,1], Charlton J. Chen, Xiaodong Yang, Chad A. Husko, and Chee Wei Wong

*Optical Nanostructures Laboratory, Columbia University, New York, NY 10027 USA*

[1]*Center of Mathematics Research, Bell Laboratories, Murray Hill, New Jersey 07974 USA*

We propose asymmetric resonant cavities with rational caustics and experimentally demonstrate interior whispering gallery modes in monolithic silicon mesoscopic microcavities. These microcavities demonstrate unique robustness of cavity quality factor ($Q$) against roughness Rayleigh scattering. Angle-resolved tapered fiber measurements and near-field images observe distinct resonant families and asymmetric emission from interior whispering gallery modes, which can be used for microcavity laser and CQED applications.

---

[*] Corresponding authors: jg2499@columbia.edu, pheider@research.bell-labs.com

Photon confinement and processes in microcavities [1-3] are critical for a vast span of fundamental studies and applications, ranging from novel zero-threshold microcavity lasers; nonlinear frequency generation [4-5]; dynamic filters and memory for communications; to interactions of atoms with cavity modes in both strong and weak coupling regimes in cavity quantum electrodynamics (QED) [6-7]. Microcavities, such as 2-dimensional disk-like resonant structures, with a single emitter possess remarkable possibilities towards efficient single photon sources for quantum communications and computing. Characterized by the cavity quality factor ($Q$; photon lifetime) and modal volume ($V$; field intensity per photon), silica microdisk-like cavities have achieved remarkable $Q$ up to $\sim 10^8$ with $V \sim 450(\lambda/n)^3$ [2]. A desirable characteristic of these planar whispering gallery mode resonators is that of directional emission, without significant $Q$-spoiling from the loss of rotational symmetry [8-11]. Different approaches have been employed by deforming the circular shape into quadruple [8], stadium [12], spiral, and line defect [13]. All these approaches need to find an optimized shape deformation parameter to support special cavity modes which possess high $Q$ as well as emission directionality. Most of the semiconductor or polymer cavities show directional lasing in the far field distribution and indeed further the advances in cavity QED and ultra-low threshold lasers. These asymmetric resonant cavities (ARC) also have potential applications in the study of disorder and localization in mesoscopic systems as well as, in the diffusive limit, quantum chaos [14] in the direct measurements of the eigenfunctions and eigenvalues of the billiard geometries. In this Letter, we show our unique approach to design ARCs with interior whispering gallery modes and characterize their directional radiation in monolithic silicon microcavities using angle-resolved tapered fiber coupling techniques and near-field images.

The asymmetric resonant cavities we construct here possess whispering gallery modes spatially located deep inside the resonator [15]. These cavities have the special property that one can inscribe into the boundary a one-parametric family of *p*-periodic orbits of the associated billiard-map. The shapes of such resonators are constructed numerically. The boundary can be computed as a solution of a nonholonomic dynamical system [15]. Here we demonstrate a class of resonators with a family of 4-periodic orbits in the numerical simulations and experiments. The involute of all these 4-periodic orbits forms the rational caustic of the shape. A caustic has the property that every tangent trajectory stays tangent after reflection at the boundary. A rational caustic defines an envelope of rays as a curve of concentrated light. In the surface of section (SOS) plot which represents the classical ray motion in phase space [16], a rational caustic corresponds to an invariant curve consisting of periodic orbits. Under slight perturbation this curve breaks up and a near integrable region appears with the usual periodic orbits surrounded in phase space by elliptic islands, invariant curves and chaotic regions. This region can be seen in the SOS of Figure 1(a) which is called the interior whispering gallery region (IWG). A trajectory starting in this region is confined to this region after many bounces at the boundary. One would expect a family of quasimodes or resonances localized near the original rational caustic. Light will mainly escape at the location where the near integrable region resulting from the rational caustic is closest to the line of total internal reflection. Emission is expected to have strong directionality where half of the total radiation comes from positions according to $\varphi = 90º$, $\theta \sim 45º$ or $135º$ in SOS map with lateral divergence angle of $\sim 11º$ in the azimuthal direction.

Analogous to whispering gallery modes one can find a family of modes, which are supported near the rational caustic. The mode in Figure 1(b) is supported along the rational caustic and has $Q \approx 1.8 \times 10^4$. There are also families of whispering gallery modes (WGM) in this kind of cavity, which are shown in Figure 1(c), located at the boundary of the 2D disk-like structure and having $Q \approx 3.0 \times 10^8$. We calculate TM modes and $Q$ factors by reducing the 3D Maxwell equations to a 2D resonance problem and solve this with a boundary integral method [17]. The size and geometry of the ARC in the simulations are the same with our fabricated ARC in experiment. Our analysis also shows that these IWG modes have linear modal volumes typically ~ 10 $(\lambda/n)^3$ and nonlinear modal volumes [5] ~ 50 $(\lambda/n)^3$, and the numbers for our WGM modes are ~15 $(\lambda/n)^3$ and 37 $(\lambda/n)^3$. In order to study the intrinsic loss mechanisms in our ARCs, we introduce slight perturbations to the cavity shape to serve as sidewall disorder roughness (in the actual fabrication) and investigate numerically these effects on $Q$s for both the fundamental WGMs and IWG modes. The intrinsic loss mechanisms in microcavities include radiation loss, scattering loss, material absorption and surface absorption [18]. For our modal volumes, surface and material absorption-limited $Q$s are typically on the order of $10^6$ [18], and the fabrication and experiments of our ARC later also show that effects of surface and material absorptions here are negligible because the radiation and scattering loss are dominant in our ARCs. The radiation $Q$ of WGMs is theoretically ~$10^8$ and surface scattering is the only mechanism which will predominantly dominate the total loss. However, the radiation $Q$ of the IWG modes is around $10^4$ because we intentionally construct the asymmetric shape to obtain directional emission, so radiation is the major loss mechanism for small edge roughness. But then scattering loss will become comparable to radiation loss when the roughness is very large. Figure 1(d) shows an exponential drop of $Q$ for the WGM with increasing edge roughness, whereas $Q$ for the IWG is only slightly affected by a perturbation larger than 15 nm. Even when the roughness is larger than 20nm, we still get a less significant spoiling of $Q$ by boundary imperfection for IWG modes than for WGMs because IWG modes are mostly concentrated along the rational caustic and away from the boundary. This indicates that the IWG modes not only possess the property of directional emission as designed, but also are not as sensitive as WGM to Rayleigh scattering from edge roughness.

Having designed this special class of asymmetric resonant cavities, we fabricate the resonators from silicon-on-insulator (SOI) wafers consisting of a 200 nm thick single-crystal Si layer on top of a 3 um $SiO_2$ cladding layer. The fabrication process starts with spin-coating of 200 nm thick 495K A6 PMMA (polymethylmethacrylate) on top of the SOI wafer. The designed pattern is written on the PMMA by electron-beam lithography. The exposed PMMA is developed in a solution of MIBK: IPA (1:3) for 55 seconds. A 30 nm chrome mask is deposited on the top of SOI wafer by thermal evaporation, and an inductively coupled-plasma (ICP) reactive-ion etch with $SF_6$:$C_4F_8$ gas chemistry then transfers the mask into the top Si layer. After the chrome mask is removed, the $SiO_2$ layer is selective etched in buffered hydrofluoric acid to create the suspended microdisk-like structures [19]. An example scanning electron micrograph (SEM) of a microfabricated resonator is illustrated in Figure 2(a), with an estimated line-edge roughness of 17 nm.

To characterize the silicon ARC, a tapered optical fiber setup such as described in Ref. [20-21] is used. By keeping an adiabatic taper profile, a SMF-28 fiber is pulled into a tapered fiber with ~ 1.1 μm waist diameter and negligible loss. The tapered fiber is then curved [20] to provide coupling and contact with the edge side of the ARC. The radius of curvature of the curved tapered fiber is ~150 um. Light from an amplified spontaneous emissions (ASE) source is passed through an in-line fiber polarizer and a polarization controller, which can rotate polarization of the light into TE or TM. This polarized light is then evanescently coupled into cavity through tapered fiber. The transmission spectrum is observed using an optical spectrum analyzer (OSA). The imaging system consists of a 50x long working distance objective lens and an 8X telescope, which provides us to monitor the fiber coupling position and collect near-infrared radiation images. Figure 2(b) shows the taper transmission spectra, illustrating clear resonant cavity modes. A manual rotational stage is used to choose different coupling positions, and the coupling depth differences at each position reflect the asymmetric property of this resonator. Positions (3) and (4) show better coupling efficiency due to the strongest mode overlap in the evanescent field with the curved tapered fiber. At positions (1) and (2), the coupling to the resonant modes does not show strongly and asymmetric Fano lineshapes [22] also appear which we suspect are due to an interference pathway between a direct background (weak tapered fiber cavity) and indirect resonant (ARC) pathway. Figure 2(c) shows both TE-like (dominant $E$-field in plane of 2D resonator) and TM-like polarizations at position (3), where a stronger TM-like coupling is observed. We select the TM-like mode in most of our measurements due to its stronger coupling and significantly reduced sensitivity to cavity edge roughness [19].

In order to examine the radiation directionality, we perform our tapered fiber coupling measurements along different coupling angles. We keep the same coupling position of the tapered-fiber on the edge of the cavity, and change the relative angle between the tapered-fiber and the resonator by rotating the stage. The transmission plot Figure 3(a) shows that two distinct families of TM modes with free spectral range ~ 11 nm are strongly guided. To identify these families, we examined the coupling depth versus $\theta$, the coupling angle between the resonator and the fiber (as defined in Figure 2(a)), at position (3) for these two modes. Three curves in Figure 3(b) shows the experimental measurement results. There exists a critical angle $\theta \sim 30º$ for the IWG modes where the coupling depth is the largest. Best excitation and output for the IWG modes occurs here. For the WGMs, no peaks appear and the coupling depth continuously decreases as the coupling angle increases, which shows that the mode overlap between the WGMs and fiber mode decreases when the tapered fiber is far away from the tangent position. This gives a clear indication that we can distinguish IWG modes and WGMs from different coupling-angle dependences. The strongest coupling strength tells us the cavity mode radiation is strongest in this direction, which matches consistently with our numerical predictions of directional emission. Note that the WGM family typically shows a larger $Q$ of 8000 than in IWG family ($Q$ of 2800) due to the increased leakage of IWG modes from directional emission. We also performed control measurements of WGMs in circular microdisk as a comparison. The coupling depth decreases exponentially as we expect, and the slight discrepancy between the WGMs in our ARC and WGMs in circular microdisk is due to the asymmetric shape of our resonator. Inset of Figure 3(b) shows the numerically calculated mode overlaps, which are proportional to the coupling depths, between different cavity modes and the fiber mode. We observed consistently the strongest mode overlap between the IWG modes and the fiber mode when $\theta \sim 45°$, which matches our experiment data and design purpose. We attribute the off-set between the measured and calculated peak positions primarily to the small deviation between our model and the actual shape, and uncertainties in the exact angular determination.

To investigate individual modes distribution in ARC, we next examine the near-field modal scattering with a tunable laser and a ThermCAM Merlin near-infrared camera, as shown in Figure 3(c). When it is tuned on-resonance with the IWG mode in Figure 3(c2), *only* three bright scattering regions in the near-field images are consistently observed. This corresponds directly with three regions of high intensity in the IWG mode as shown in Figure 1(b), and also corresponds to three peaks within the interior whispering galley region in the SOS map in Figure 1(a). We believe that these three scattering regions represent where the directional emission of the ARC happens. Figure 3(c1) shows a WGM mode when tuned on-resonance. We can see that the escape route of WGMs is totally different with IWG modes. Scattering from the entire boundary for WGMs can be observed, which also confirms our simulation result that scattering loss of WGM modes due to edge roughness is dominated while IWG modes is not that sensitive to the edge roughness. From previous angle-resolved coupling measurements, we also note that at this coupling position (position 3 and $\theta = 0$) the WGM mode couples better to the tapered fiber, compared to the IWG mode. This is the reason why Figure 3(c1) looks brighter than Figure 3(c2) under the condition of the same input power from the tunable laser. As a reference, a near-infrared image with off-resonance condition is also illustrated in Figure 3(c3), in which only the fiber mode is observable.

To characterize the quality factor of IWG modes, feedback-controlled piezoelectric stages are used to control the lateral taper-cavity separation $g$, which is illustrated in the inset of Figure 4(b). The transmission spectrum in Figure 4(a) shows an initial $Q$ estimate of 2800 for the mode at 1543.44 nm when $g = 0$. This is a lower bound for cavity $Q$ due to the taper's loading effects. In Figure 4(b), these loading effects diminish as the taper is positioned further and further above the cavity, until a regime is reached where $Q$ does not change. This asymptotic behavior gives us the estimate an intrinsic cavity $Q \sim 6000$, and also implies that we are working in the under-coupled region. The measured $Q$ deviates from the numerical predictions due to increased surface scattering in the vertical direction (not captured in simulations).

In summary, we have constructed numerical methods to design remarkable resonators with 4-periodic rational caustics and, in particular, interior whispering gallery modes with high cavity $Q$s for strong light-matter interactions. We have observed experimentally the radiation directionality of the optical modes with $Q$ value of ~ 6000 in the asymmetric resonant cavities through tapered fiber coupling techniques. The $Q$ values of these modes are not spoiled by slight edge boundary imperfection, which is an advantage for application and proves the usefulness of resonators with rational caustics together with the directionality. Our design and characterization of interior whispering gallery modes in silicon asymmetric resonant cavities opens the door to engineer the asymmetric cavity shape and achieve its application in microcavity-based QED and lasing with directional emission in the future.

The authors acknowledge support by the Columbia Initiative in Nanophotonics, NSF ECS 0622069, the New York State Office of Science, Technology and Academic Research. P.H. is supported by a Feodor Lynen scholarship of the Humboldt-Foundation. X.Y. is supported by an Intel fellowship. We thank Y. Baryshnikov for helpful discussions, K. Srinivasan and O. J. Painter for kind suggestions on the tapered fiber preparation, and B. Lee and K. Bergman for loaning the ThermCAM Merlin for near-infrared imaging.

**Figure Captions:**

Fig. 1 (color online). (a) Poincaré surface of section plot for ARC used in the experiment. Horizontal axis ϕ represents the intercept point of the ray and the cavity boundary, ranging from 0 to 2π according to the coordinate system shown in the inset. The vertical axis θ represents the incident angle between the ray and the tangent line of the boundary [16]. The blue solid line corresponds to the total internal reflection. (b) Interior whispering gallery mode supported near the rational caustic and (c) the fundamental WGM mode. The ARC boundary is shown in black as a reference. (d) Numerical simulations illustrate that $Q$ of WGMs (squares) drops exponentially with cavity edge roughness, while it stays nearly constant for IWG modes (circles). For the edge roughness, we use r_perturbed (ϕ) = r_unperturbed (ϕ) + roughness * sin (100 * ϕ) along the boundary as a perturbation.

Fig. 2 (color online). (a) SEM of fabricated silicon ARC. Dotted lines (--) show tapered fiber spatial probe positions [(1) – (4)]. Scale bar: 10 μm. Inset: isometric side view of suspended silicon ARC on oxide pedestal. (b) Normalized taper transmissions when the taper touches the resonator side in tangential direction at different positions [(1) - (4)]. (c) TE-like and TM-like modes can be selected by polarization controller, and we optimize TM-like in all the experiments.

Fig 3 (color online). (a) Normalized transmission response of ARC when the taper is positioned along different coupling directions at position (3). (b) Experimental results (dots for WGM modes, squares for IWG modes and triangles for WGM modes in circular microdisk) for coupling depth versus $θ$. Inset: Numerical (red curve for WGM and green curve for IWG mode) results for coupling depth versus $θ$. (c) Near-infrared images of ARC at different frequencies: (c1) on-resonance with WGM family (c2) on-resonance with IWG family (c3) off-resonance, where fiber mode can also be observed.

Fig 4 (color online). (a) Transmission spectrum when fiber-cavity separation $g = 0$ and we focus on IWG modes. (b) Coupling depth (circles) and total $Q$ factor (squares) as functions of $g$. Inset: Experiment illustration for measuring the $Q$ value.

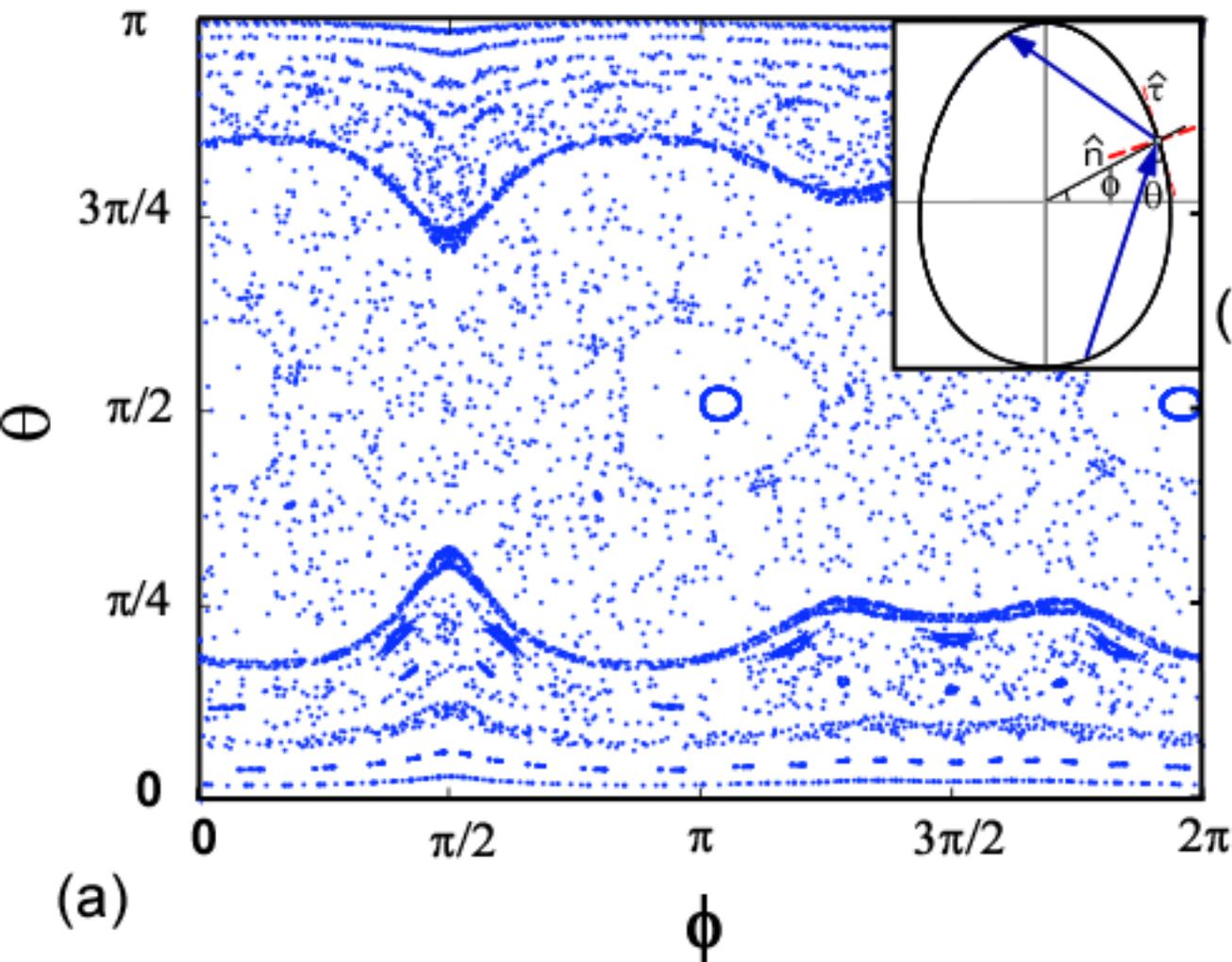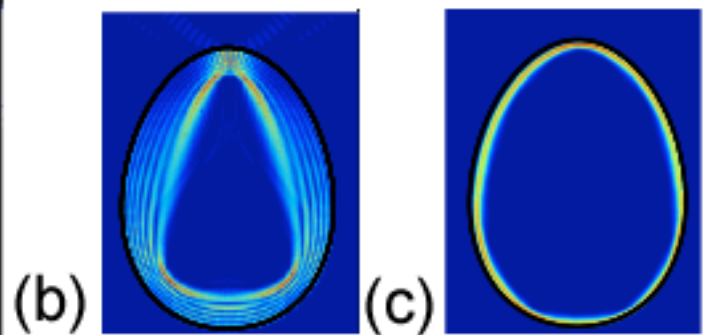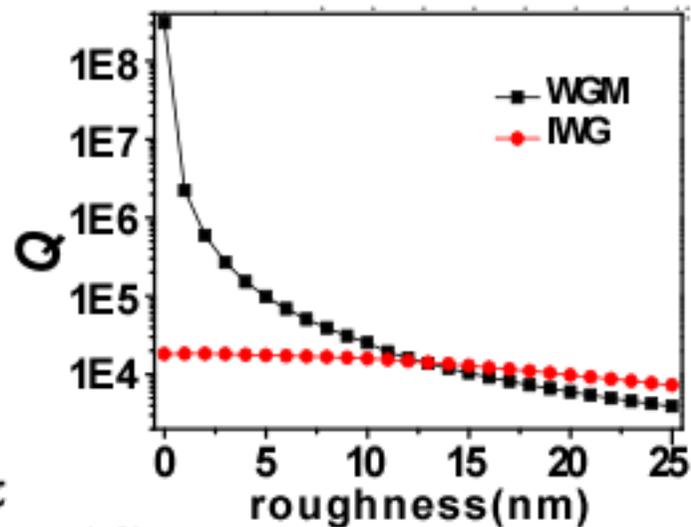

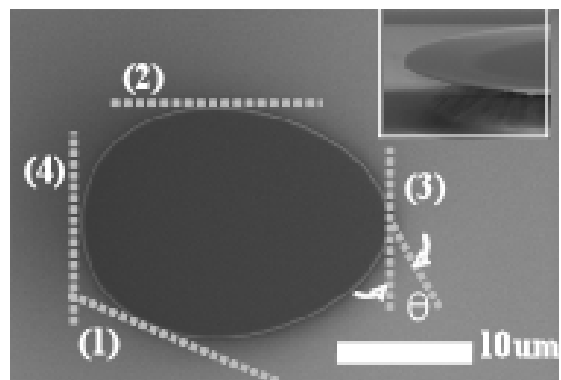
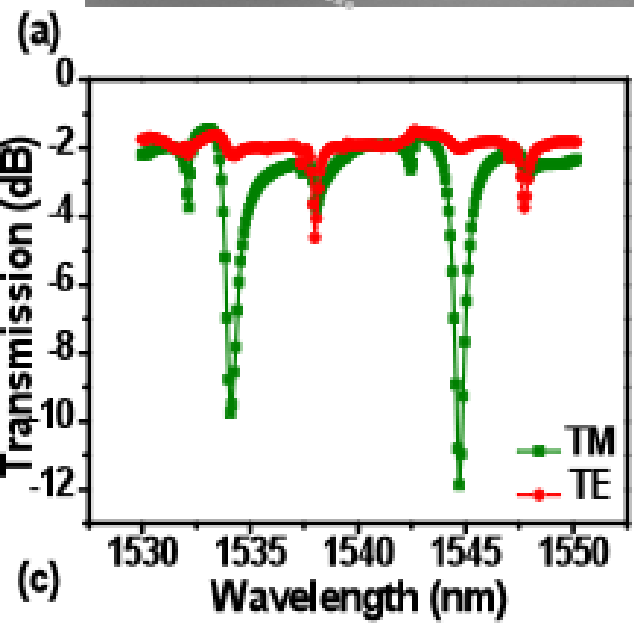
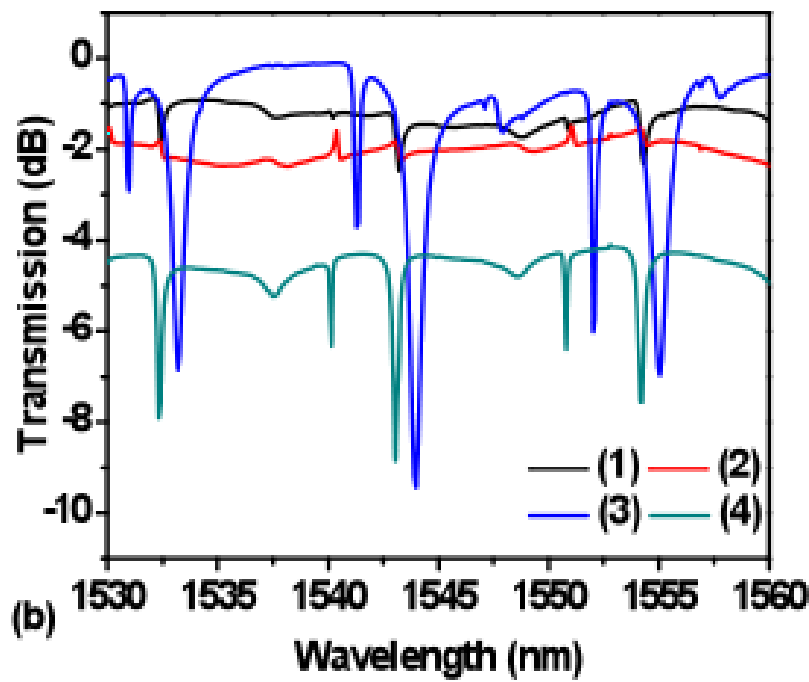

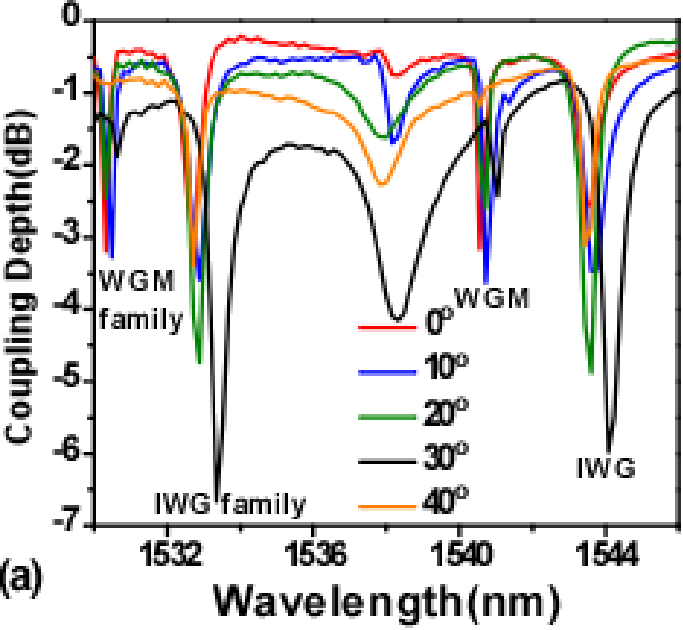
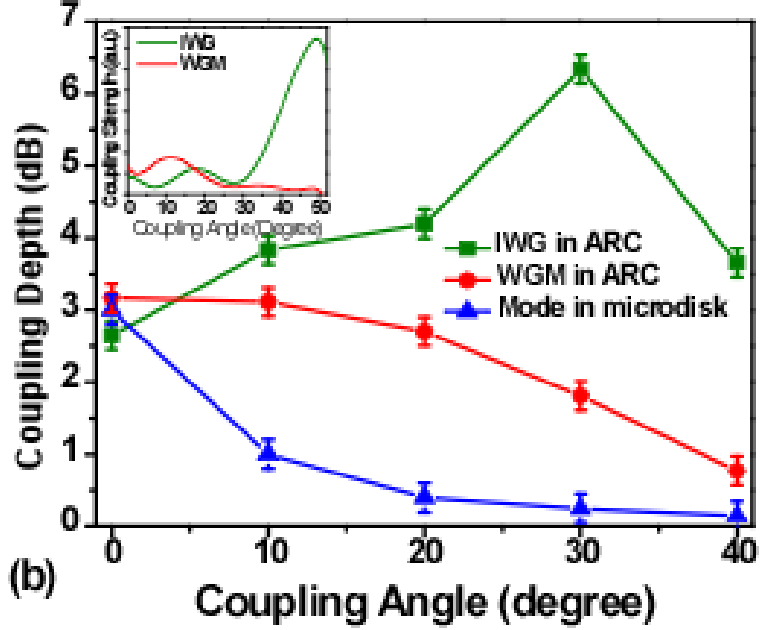
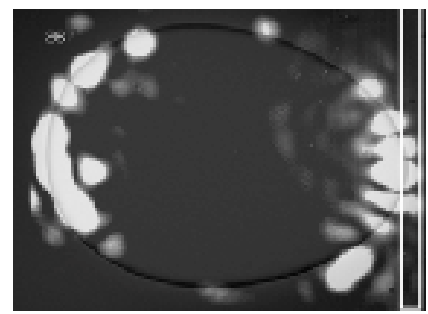
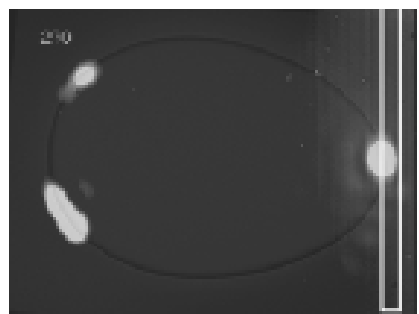
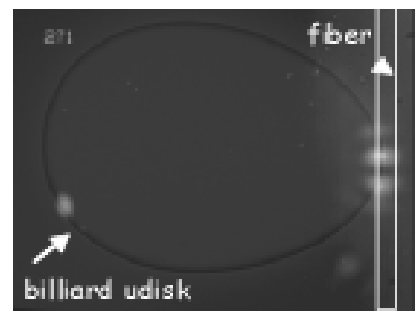

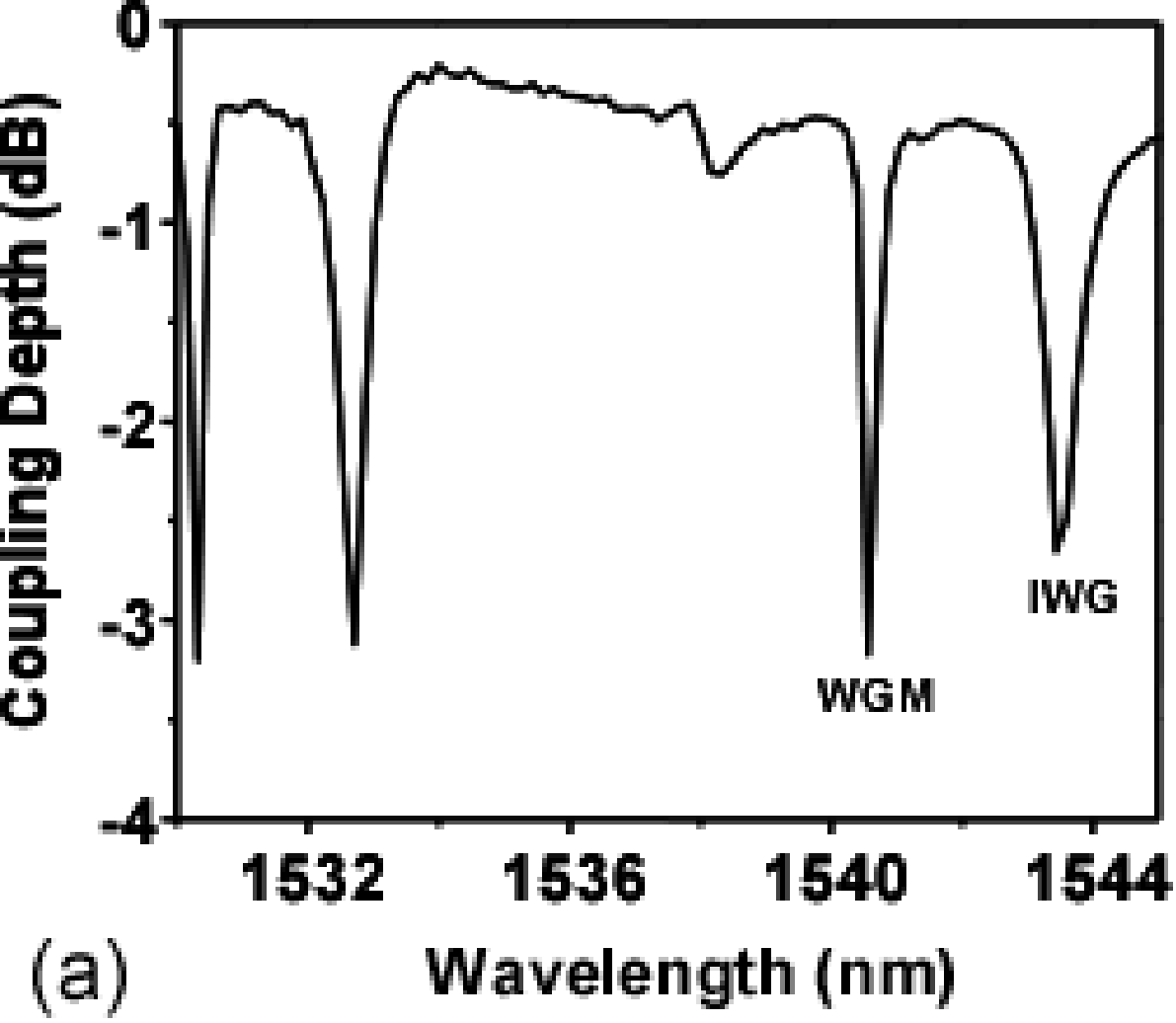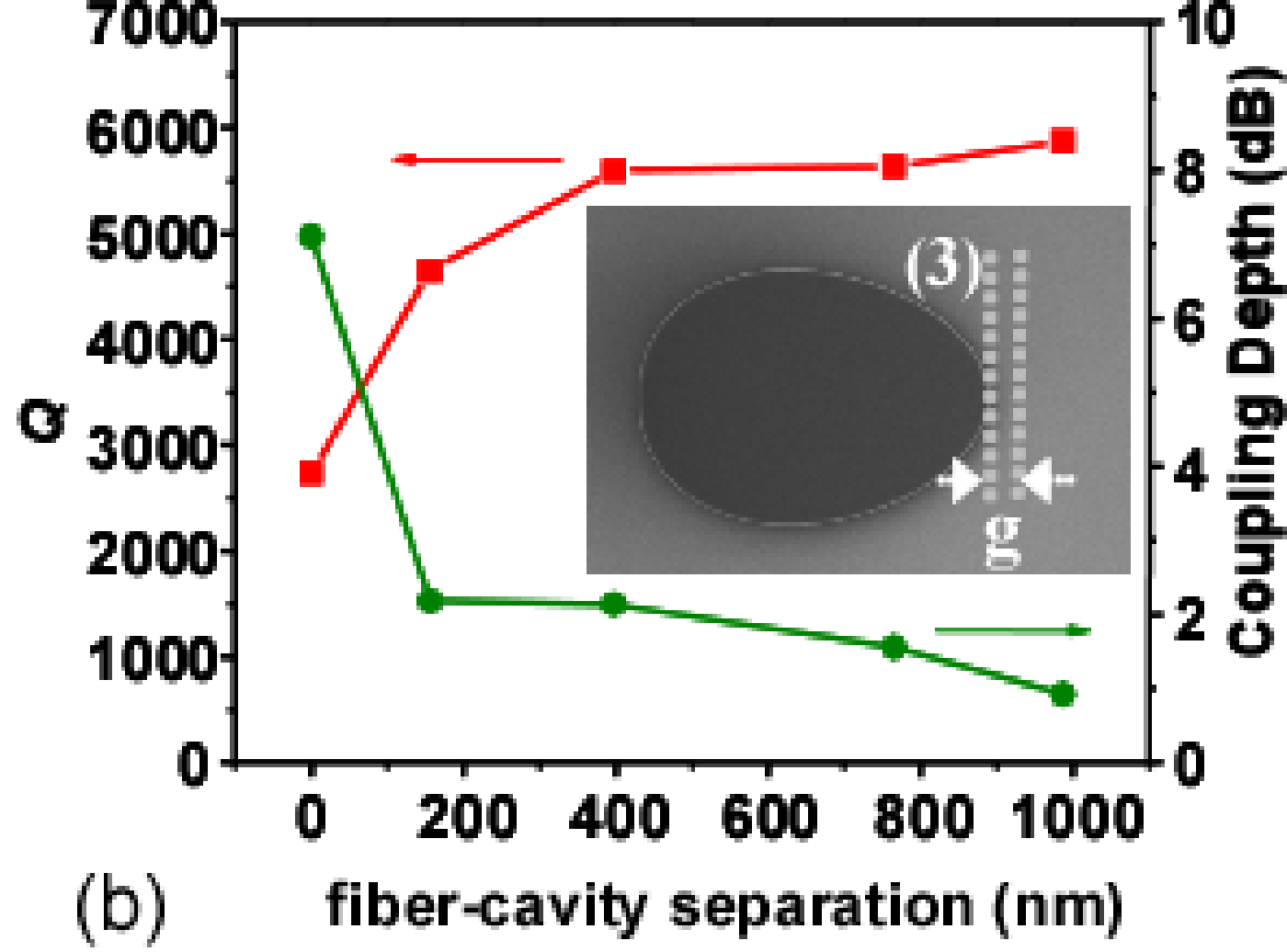